# High spatial resolution spectral imaging method for space interferometers and its application to formation-flying small satellites

**Taro Matsuo,[a,b]* Satoshi Ikari,[c] Hirotaka Kondo,[c] Sho Ishiwata,[c] Shinichi Nakasuka,[c] Tomoyasu Yamamuro [d]**

[a]Nagoya University, Graduate School of Science, Department of Basic Particles and Astrophysics, Furo-cho, Chikusa-ku, Nagoya 464-8601, Japan
[b]NASA Ames Research Center, Moffett Field, Mountain view, CA, U.S.A., 94035
[c]The University of Tokyo, Bunkyo-ku, Tokyo 113-8656, Japan
[d]Optocraft, Sagamihara, Kanagawa 252-0144, Japan

**Abstract**. Infrared space interferometers can surpass the spatial resolution limitations of single-dish space telescopes. However, stellar interferometers from space have not been realized because of technical difficulties. Two beams coming from individual satellites separated by more than a few tens of meters should precisely interfere such that the optical-path and angular differences between the two beams are reduced at the wavelength level. Herein, we propose a novel beam combiner for space interferometers that records the spectrally-resolved interferometric fringes using the densified pupil spectroscopic technique. As the detector plane is optically conjugated to a plane, on which the two beams interfere, we can directly measure the relative phase difference between the two beams. Additionally, when an object within the field of view is obtained with a modest signal-to-noise ratio, we can extract the true complex amplitude from a continuous broadband fringe (i.e., one exposure measurement), without scanning a delay line and chopping interferometry. We discovered that this spectral imaging method is validated for observing the solar system objects by simulating the reflected light from Europa with a small stellar interferometer. However, because the structure of the object spectrum may cause a systematic error in the measurement, this method may be limited in extracting the true complex amplitude for other astronomical objects. Applying this spectral imaging method to general astrophysics will facilitate further research. The beam combiner and spectral imaging method are applied to a formation-flying stellar interferometer with multiple small satellites in a Sun-synchronous orbit, named Space Experiment of InfraRed Interferometric Observation Satellite (SEIRIOS), for observation of the solar system objects in visible and near-infrared. We present an overview of SEIRIOS and the optimized optical design for a limited-volume spacecraft.

*First Author, E-mail: matsuo@u.phys.nagoya-u.ac.jp

## 1. Introduction

Infrared astronomy sheds light on the history of the baryon universe, from the formation of the first stars and galaxies to the birth of life on Earth. However, the spatial resolutions of the current mid-to-far-infrared (10 – 300 $\mu m$) observatories are significantly limited because of the longer wavelengths. Radio observatories[1,2,3] can break their diffraction limits owing to interferometric techniques. The stellar interferometer is the most promising path toward implementing milli-arcsecond spatial resolution at infrared wavelengths. Several optical and infrared ground-based observatories such as the Keck interferometer[4], the Very Large Telescope Interferometer (VLTI)[5], the Large Binocular Telescope Interferometer (LBTI)[6], the Center for High Angular Resolution Astronomy (CHARA)[7], and the Navy Precision Optical Interferometer[8], have realized stellar interferometry by combining the light from more than two ground-based telescopes. While radio interferometers detect photons at each telescope before combining them, infrared interferometers must directly combine the photons from two or more telescopes on a detector plane before detecting them. As a result, designing optical and infrared interferometers is much more difficult than radio interferometers because the optical path difference between the two beams is smaller than the observing wavelength under atmospheric turbulence. The number of telescopes comprising most of a stellar interferometer is less compared to radio telescopes, such as the Atacama Large Millimeter/Submillimeter Array (ALMA)[1] and the Square Kilometer Array (SKA)[2]. Consequently, optical stellar interferometers have lower $U$-$V$ coverage than ALMA and SKA. Thus, realizing a stellar interferometer in space can enhance the imaging ability of interferometer because the $U$-$V$ plane is easily filled. In addition, the baseline length of a space interferometer in the presence of a formation flying could be comparable to those of the ground-based interferometers.

Space stellar interferometers with multiple large apertures have been advocated for a long time. The Terrestrial Planet Finder-Interferometer (TPF-I) /Darwin[9] aimed to characterize the atmospheres of potentially habitable planets orbiting nearby stars by combining imaging interferometry to the nulling technique. The Space Interferometry Mission (SIM)[10] was planned to detect tiny astrometric signals from nearby Sun-like stars induced by Earth-mass planets. The Far-InfraRed Interferometer mission (FIRI)[11] can provide high-resolution spectral imaging in the wavelength range in which the light does not reach the ground for the first time.

However, technological developments toward future space interferometers have advanced along with mission designs and proposals. Deep Space 3[12] and the StarLight mission[13] were designed to validate the formation flying interferometry in space. Further, new potential paths toward realizing large-scale space interferometers have been investigated[14,15,17]. Owing to the rapid technological advancement in microsatellites, they have been considered the precursors' platform for future large-scale space telescopes. Formation-flying interferometry with three small satellites in Low Earth Orbit has been investigated in terms of the feasibility and deemed feasible with the current technologies[17,18]. Moreover, six small spacecraft are being considered for aperture synthesis in the low-radio frequency regime[19]. A structurally connected space interferometer, proposed by Made in Space (now Redwire) and Lowell Observatory, would be formed by manufacturing a 20-meter-long boom on-orbit with 3D printer technology[20]. Related technology is Miniaturized Distributed Occulter/Telescope (mDOT)[21], while is not a space interferometer, mDOT but applies formation-flight technology to its mission. Inspired by these projects, we propose a formation-flying space interferometer comprising three small satellites under the program called Space Experiment of InfraRed Interferometric Observation Satellite (SEIRIOS). SEIRIOS will be the first-generation space interferometer under a formation flying of multiple small satellites and perform spectral imaging. The baseline of SEIRIOS changes from 10 to 100 $m$, and the observation wavelength ranges from 0.6 to 1.7 $\mu m$. However, despite the rapid advances in small-satellite technology, realizing a formation-flying interferometry is difficult owing to its strict requirements. A space interferometer should interfere two beams coming from individual satellites separated by a few tens of meters at the precision of the observing wavelength. In addition, when the optical path difference between the two beams substantially fluctuates, measuring the spectrally-resolved complex visibility is difficult because the real part of the complex visibility (i.e., the cosine part) is recorded in the interferometric fringe. Interferometric chopping transforms the cosine part into a sine part by adding a phase of π/2 to one of the two beams. Double Fourier Interferometry (DFI)[22,23] applies the interference fringes obtained by changing the optical path difference to Fourier transform spectroscopy and reconstructs the spectrally-resolved complex visibility. Although the complex visibility can be obtained from one interference fringe when DFI is applied to the Fizeau-beam combiner, the resolving power is limited by the number of the fringe patterns on the detector plane[24].

In this study, we propose a novel beam combiner for space interferometers. We develop a densified pupil spectrograph[25] originally proposed for high-precision transit spectroscopy. This beam combiner measures the complex visibility (i.e., the cosine and sine parts) from the interference fringe taken at one exposure. In addition, the recorded interference fringe provides information on the relative phase difference between the two beams, such as the optical path difference and relative tilt. From continuous broadband fringe measurements, we can reconstruct the true complex amplitude when an astronomical object within the field of view is recorded with a modest signal-to-noise ratio. The phase difference between the two beams is determined by the angular position of the astronomical object and the optical path difference between the two beams generated in the interferometer. As the phase of the true complex visibility includes only the former (i.e., the angular position of the astronomical object), the images reconstructed from the true visibilities are non-ambigous. We note that the true phase can also be obtained by the closure phase technique[26], which uses more than three telescopes in general stellar interferometers. Using the proposed beam combiner, which will be employed in SEIRIOS, we can reconstruct the true complex visibility with only two collectors. The proposed method could be applied not only to the stellar interferometers using small spacecraft but also to future space interferometers. The Large

Interferometer For Exoplanets (LIFE)[27] has been started as a revival of TPF-I/Darwin. Future space interferometers for general astrophysics are also demanded across the electromagnetic spectrum in NASA's Astrophysics Roadmap[28].

Using the proposed beam combiner, we reconstruct the true complex amplitude from one exposure measurement. The beam combiner is then applied to SEIRIOS as an example. Section 2 overviews the proposed beam combiner and describes the optimized spectral-imaging method for the beam combiner. Section 3 analytically describes the measurement and calibration of the interferometric fringes with the proposed beam combiner, and Section 4 summarizes the requirements for obtaining the complex amplitudes. Finally, Section 5 overviews the SEIRIOS project and applies the beam combiner to SEIRIOS.

## 2. Concept

In this section, we first describe a new stellar interferometer concept that employs the densified pupil spectrograph as a Michelson-type beam combiner. The novel beam combiner allows the measurements of the complex amplitude and relaxes the requirements for generating the interference fringes. Then, we develop a procedure that measures the artificial phase errors and reconstructs the true complex amplitudes from one exposure measurement.

### 2.1. New type of beam combiner

The proposed beam combiner, based on the densified pupil spectrograph, is named "densified pupil spectrograph interferometer (DPSI)" and enables complex visibility measurements. Figure 1 is a conceptual diagram of the DPSI. The densified pupil spectrograph originally proposed for enabling high-precision spectrophotometry from space generates the spectra of the densified sub-pupils on the detector plane optically conjugated to the primary mirror. The densified pupil spectrograph divides an entrance pupil with a pupil slicer, and each divided sub-pupil is densified with two concave mirrors. Because the size of each sub-pupil is comparable to one hundred times the wavelength, each densified sub-pupil is spread as a point source. The spectra of the densified sub-pupils are formed on the science detector after the beam passes through a dispersive element. Since the detector of the spectrograph is optically conjugated to the primary mirror, the spectrum is in principle insensitive to the telescope pointing jitter.

When the densified pupil spectrograph design is applied as the beam combiner of a stellar interferometer, the detector plane for recording the interference fringes is optically conjugated to the interference plane of the two beams (i.e., the beam splitter). DPSI provides three main advantages over conventional beam combiners. First, the beam combiner enables a direct measurement of the relative phase difference between the two beams on the pupil plane, unless the angular diameter of the target object exceeds the spatial resolution of each beam. Consequently, the relative tilt can be compensated even after recording the interference fringes (see Section 2.2). In addition, if the relative tilt slowly changes, it can be compensated in real-time. The second advantage is that reconstruct the complex visibility (i.e., cosine and sine components) can be reconstructed from one exposure measurement without delay-line changing and chopping interferometry. As the formed interference fringe is optically conjugated to the interference plane, the sine component of the complex visibility can be obtained by adding a phase of $\pi/2$ to half of the entrance pupil of one of the two beams. The cosine and sine components of the complex visibility are simultaneously measured from the same exposure measurement. The third advantage is mitigation of the conditions for generating the interference fringe.

To form the interference fringe, the following three conditions must be satisfied. (1) Two beams are overlapped on the interference plane, (2) the optical path difference between the two beams is less than the coherence length, and (3) the interference fringe is not smoothed by fluctuations of the relative angle and the optical path difference during each exposure time. The first condition mitigates the requirement on the pointing jitter of the stellar interferometer. Because the densified pupil spectrograph has medium spectral-resolving power, it provides a relatively long coherence length. Further, the second condition is mitigated by the DPSI compared to the conventional beam combiner. The third condition is imposed on all the beam combiners, including DPSI. An additional requirement is the accurate phase shift of $\pi/2$, which minimizes the systematic error in the complex visibility measurement. In Section 4, we analytically derive the four conditions for accurately measuring the complex visibility and quantitatively evaluate them in a SEIRIOS application of the DPSI.

Although the densified pupil spectrograph requires an additional optical component compared to conventional ones, the latest design can be greatly simplified. The pupil is subdivided using a compact beam slicer named the Bowen-Walraven type image slicer[29,30]. As a secondary advantage, the pupil division maintains the compactness of the highly dispersed spectrograph design compact[31]. This simple and compact optical design is applicable not only to general large satellites but also to limited-volume spacecraft. In Section 5, we apply the new beam combiner to SEIRIOS and disucss its optical design.

*2.2. Measurement of true complex amplitude*

The complex amplitude of each pixel and each spectral element is reconstructed from one exposure measurement. However, the reconstructed complex amplitudes are affected by the relative tilt and optical path difference between the two beams. Here, we develop a spectral imaging method that reconstructs the true complex visibility from the interference fringes obtained by DPSI. Figure 2 summarizes the procedure from data acquisition to derive the true complex amplitude. Focusing on the first advantage of DPSI introduced in Section 2.1, we first measure the relative tilt between the two beams from the phase distribution of the reconstructed complex visibility (Process (2) of Figure 2). Then, we remove the effect of the relative tilt on the complex amplitude (Process (3) of Figure 2). Next, we co-add the calibrated complex amplitudes over the $n_{pix} \times n_{pix}$ pixels of each spectral element, where $n_{pix} \times n_{pix}$ is the assumed sampling area of the pupil for each spectral element (Process (4) of Figure 2). Finally, the complex amplitudes with the optical path difference generated by the pointing error and the internal asymmetric optical path length are obtained.

The DPSI can also calibrate the artificial optical-path difference between the two beams. A continuous observing bandwidth is obtained from space, so its intensity distribution on the sky can be reconstructed by Fourier transformation of the coadded complex amplitudes along the spectral direction[32] (Process (5) of Figure 2). Because the position of the reconstructed astronomical object reflects the artificial phase component, we can calibrate the complex visibility using the reconstructed image. By measuring the true complex visibility at each point on the *U-V* plane, we can minimize the uncertainty in the image reconstruction based on the Van Cittert-Zernike theorem. Thus, DPSI paves the way to perform for high-spatial resolution spectral imaging even under the unstable observing conditions of formation-flying interferometry with multiple small satellites. This spectral imaging method is analytically described in Section 3.

**3. Spectral Imaging Method**

This section analytically reconstructs the spectrally resolved complex visibility for high-spatial resolution spectral imaging. We first derive the measurement value recorded by the densified pupil spectrograph under the optical path and angular differences between two beams. We then introduce a method to reconstruct the spectrally resolved true complex visibility from the described measurement values without delay-line scanning and chopping interferometry.

*3.1. Interference fringe*

Under the condition that the two daughter satellites are ideally positioned, the coordinate system and parameters on the entrance aperture of SEIRIOS are shown in Figure 3. The origin of the coordinate system is set as the center position of the line connecting the two daughter satellites. For simplicity, each aperture is set to a square of length *D*, and the position vectors of the daughter satellites 1 and 2 are $-\frac{\vec{B}}{2}$ and $+\frac{\vec{B}}{2}$, respectively. The coordinate system on the interference plane, optically conjugated to the detector plane, is set to $(x, y)$. The baseline vector of $\vec{B}$ is aligned to the *x* axis under the ideal condition shown in Figure 3. The changes of the optical path lengths caused by the displacements of the daughter satellites from the ideal positions are set to $c_1$ and $c_2$, respectively. Notably, the displacement of each daughter satellite along the *y* axis has no impact on the optical path difference between the two beams. However, to ensure overlap of both beams on the interference plane, the displacement along the *y* axis should be less than the size of the entrance pupil. The angles around the $x$ and $y$ axes are defined as $\beta$ and $\alpha$, respectively. We note that the angle around the $z$ axis is not defined because the rotation of the plane mirror does not affect the fringe measurement. The position vector from the origin of the coordinate system to the astronomical object is $\vec{\theta}$. The incident electric field to each aperture is $E_0$.

Based on the above definition of the coordinate system and parameters, we formulate the complex amplitude obtained by SEIRIOS. The electric fields passing through the entrance apertures 1 and 2 and arriving at the interference plane (i.e., a polarizing beam splitter) are described as

$$E_1(x, y, \vec{\theta}) = E_0(\vec{\theta}, \lambda) e^{i\frac{2\pi}{\lambda}(-\frac{\vec{B}\cdot\vec{\theta}}{2} + \alpha_1 x + \beta_1 y + c_1)}, \tag{1}$$

$$E_2(x, y, \vec{\theta}) = E_0(\vec{\theta}, \lambda) e^{i\frac{2\pi}{\lambda}(\frac{\vec{B}\cdot\vec{\theta}}{2} + \alpha_2 x + \beta_2 y + c_2)}, \tag{2}$$

where $\alpha_i$ ($\beta_i$) is the angle of the $i$-th aperture around the $y$ ($x$) axis from the ideal condition. The electric field formed on the science detector is

$$E(x, y, \vec{\theta}) = E_1(x, y, \vec{\theta}) + E_2(x, y, \vec{\theta})$$

$$= E_0(\vec{\theta}, \lambda) \left\{ e^{i\frac{2\pi}{\lambda}(-\frac{\vec{B}\cdot\vec{\theta}}{2} + \alpha_1 x + \beta_1 y + c_1)} + e^{i\frac{2\pi}{\lambda}(\frac{\vec{B}\cdot\vec{\theta}}{2} + \alpha_2 x + \beta_2 y + c_2)} \right\}. \tag{3}$$

Since the science detector for recording the interference fringes is optically conjugated to the interference plane by the densified pupil spectrograph, the coordinate system of the science detector is also described by the same coordinate system. The following interference fringe is observed on the science detector:

$$I(x, y, \vec{\theta}) = |E(x, y, \vec{\theta})|^2$$

$$= |E_0(\vec{\theta}, \lambda)|^2 (2 + e^{i\frac{2\pi}{\lambda}(\vec{B}\cdot\vec{\theta} + (\alpha_2 - \alpha_1)x + (\beta_2 - \beta_1)y + (c_1 - c_2))} + c.c.), \tag{4}$$

where $c.c.$ represents the complex conjugate of the second term on the right-hand side. The second and third terms are called interference terms, which include the complex visibility. By blocking one of the two beams before the interference plane, we acquire the intensity of the single aperture, $|E_0(\vec{\theta},\lambda)|^2$, instead of the interference fringe, and the interference term is separated from the first term by subtracting $|E_0(\vec{\theta},\lambda)|^2$ from the observed interference fringe:

$$I(x,y,\vec{\theta}) - 2|E_0(x,y,\vec{\theta})|^2$$

$$= 2|E_0(\vec{\theta},\lambda)|^2 \cos\left(\frac{2\pi}{\lambda}(\vec{B}\cdot\vec{\theta} + (\alpha_2 - \alpha_1)x + (\beta_2 - \beta_1)y + (c_1 - c_2))\right)$$

$$= V(\vec{\theta},\lambda) e^{i\frac{2\pi}{\lambda}((\alpha_2-\alpha_1)x+(\beta_2-\beta_1)y+(c_1-c_2))} + c.c., \tag{5}$$

where $V(\vec{\theta},\lambda)$ shows the true complex amplitude (without any systematic effects), which is important for performing the spectral imaging:

$$V(\vec{\theta},\lambda) = |E_0(\vec{\theta},\lambda)|^2 e^{i\frac{2\pi}{\lambda}\vec{B}\cdot\vec{\theta}}. \tag{6}$$

Equation (5) shows that the recorded interference fringe is affected by the optical path and angular differences between the two beams. In addition, even if these errors are successfully calibrated, only the cosine component of the complex amplitude is extracted from the recorded interference fringe. In this case, the stellar interferometer is sensitive to only the symmetric components of the source plane.

*3.2. Relative complex visibility*

Both the sine and cosine components are necessary for reconstructing the complex visibility. We can measure the sine component by adding a quarter of the wavelength to the phase for one of the two beams[33]. This interferometric chopping is also applied to the direct detection of a faint companion orbiting the bright star[34]. When a quarter of the wavelength is added to the phase of Aperture 2, Equations (2) is rewritten as

$$E_2'(x,y,\vec{\theta}) = E_0(\vec{\theta},\lambda) e^{i\frac{2\pi}{\lambda}\left(\frac{\vec{B}\cdot\vec{\theta}}{2}+\alpha_2 x+\beta_2 y+c_2-\frac{1}{4}\lambda\right)}. \tag{7}$$

In this case, the extracted interference term corresponding to Equation (5) becomes

$$I'(x, y, \vec{\theta}) - 2|E_0(x, y, \vec{\theta})|^2$$

$$= 2|E_0(\vec{\theta}, \lambda)|^2 \sin\left(\frac{2\pi}{\lambda}(\vec{B} \cdot \vec{\theta} + (\alpha_2 - \alpha_1)x + (\beta_2 - \beta_1)y + (c_1 - c_2))\right)$$

$$= -iV(\vec{\theta}, \lambda)e^{i\frac{2\pi}{\lambda}((\alpha_2 - \alpha_1)x + (\beta_2 - \beta_1)y + (c_1 - c_2))} + c.c. \tag{8}$$

From Equations (5) and (8), the complex amplitude can be estimated as follows:

$$I(x, y, \vec{\theta}) - 2|E_0(x, y, \vec{\theta})|^2 + i(I'(x, y, \vec{\theta}) - 2|E_0(x, y, \vec{\theta})|^2)$$

$$= V(\vec{\theta}, \lambda)e^{i\frac{2\pi}{\lambda}((\alpha_2 - \alpha_1)x + (\beta_2 - \beta_1)y + (c_1 - c_2))}, \tag{9}$$

where we assumed that the optical path and angular differences between the two beams are constant for the measurements of the cosine and sine components. However, it is challenging to reconstruct the complex visibility with the conventional chopping technique because the optical path and angular differences between the two beams continuously change during the observation. Here, we noticed that the cosine and sine components of the complex amplitude can be simultaneously obtained by adding a quarter of the wavelength to half of the entrance aperture for one of the two beams (see Figure 4) because the densified pupil spectra are equal to those formed on the plane of the polarized beam splitter. The achromatic phase shift can be realized by a combination of two quarter-wave plates and a half-wave plate[35]. The half-wave plate is inserted between the two quarter-wave plates. First, both clockwise and counterclockwise circularly polarized beams are generated by the first quarter-wave plate. The next half-wave plate delays or advances the phases of the circularly polarized beams. The amount of the phase shift can be determined by adjusting the relative direction between the fast axes of the first quarter- and half-wave plates. Finally, the second quarter-wave plate restores the original polarization state before passing through the first quarter-wave plate. Only the phase of the P- (S-) wave is delayed (advanced) through combining the three wave plates. When the two beams interfere on the plane of the polarized beam splitter, only the P- or S-wave can be extracted into one of the two interfered beams. Thus, we obtain the complex amplitude shown in Equation (9) from one exposure measurement by adding the achromatic phase shift of a quarter of the wavelength to half of the entrance aperture for one of the two beams. We analytically derive the requirement on the phase shifter in Section 4.3.

*3.3. Calibration of the optical path and angular differences between two beams*

Although the complex amplitude can be obtained from the data taken at one exposure by the optimized achromatic phase shifter for DPSI, the obtained data include not only the complex amplitude of an astronomical object but also the phase difference between the two beams caused by the relative angle of $(\alpha_2 - \alpha_1, \beta_2 - \beta_1)$ and the optical path difference of $(c_1 - c_2)$. To extract the true complex amplitude shown in Equation (6), both the optical path and angular differences between the two beams should be compensated during data reduction. As mentioned in Section 2.2, the DPSI can directly measure the relative angle between two beams, $(\alpha_2 - \alpha_1, \beta_2 - \beta_1)$, from the phase distribution of the densified pupils formed on the detector plane. To measure the attitudes of the two daughter satellites, we employ an attitude sensor that forms the two point-spread functions on the focal plane (see Section 5). The attitude sensor is also useful when the interference fringes are not formed because the optical path difference exceeds the coherence length. The term of $e^{i\frac{2\pi}{\lambda}((\alpha_2-\alpha_1)x+(\beta_2-\beta_1)y)}$ can be estimated from the measured relative angle and $V(\vec{\theta},\lambda)e^{i\frac{2\pi}{\lambda}(c_1-c_2)}$ is obtained from Equation (9). Furthermore, since a continuous wavelength range can be covered, we can estimate the optical path difference of $(c_2 - c_1)$ from the Fourier transform of the complex amplitude, including the optical path difference along the observing wavelength[32]:

$$\int d\left(\frac{1}{\lambda}\right) V(\vec{\theta}, \lambda) e^{i\frac{2\pi}{\lambda}(c_1-c_2)} e^{-i\frac{2\pi}{\lambda}\vec{B}\cdot\vec{\theta'}}$$

$$= \int_{1/\Delta\lambda_{band}} d\left(\frac{1}{\lambda}\right) |E_0(\vec{\theta}, \lambda)|^2 e^{i\frac{2\pi}{\lambda}(\vec{B}\cdot\vec{\theta}+c_1-c_2)} e^{-i\frac{2\pi}{\lambda}\vec{B}\cdot\vec{\theta'}}$$

$$= \delta(\vec{B}\cdot\vec{\theta'} - \vec{B}\cdot\vec{\theta} - (c_1 - c_2)) * \text{sinc}\left(\frac{\pi \vec{B}\cdot\vec{\theta'}}{2\Delta\lambda_{band}}\right) * FT\{|E_0(\vec{\theta}, \lambda)|^2\}, \qquad (10)$$

where * represents a convolution operator, and $\Delta\lambda_{band}$ is the observation bandwidth. When the following condition is satisfied, $|E_0(\vec{\theta}, \lambda)|^2 = constant$ along the wavelength, the intensity distribution of the source along the baseline vector is reconstructed:

$$\int d\left(\frac{1}{\lambda}\right) V(\vec{\theta}, \lambda) e^{i\frac{2\pi}{\lambda}(c_1-c_2)} e^{i\frac{2\pi}{\lambda}\vec{B}\cdot\vec{\theta'}} = \left|E_0\left(\theta' = \theta + \frac{c_1-c_2}{|\vec{B}|}\right)\right|^2 * \text{sinc}\left(\frac{\pi \vec{B}\cdot\vec{\theta'}}{2\Delta\lambda_{band}}\right). \qquad (11)$$

The position of the source is shifted by $\frac{c_1-c_2}{|\vec{B}|}$ from its true position along the baseline vector. The internal optical path difference can be, in principle, measured with uncertainty of $\frac{\vec{B}\cdot\vec{\theta'}}{2\Delta\lambda_{band}}$ if the true position of the source is known. Even if the original position is unknown, the optical path difference can be compensated by assuming that an astronomical object on the reconstructed image is fixed at the same position. For example, the source is fixed to the center of the reconstructed image. Thus, the true complex amplitude of $V(\vec{\theta}, \lambda)$ could be extracted from one exposure measurement when the structure of the object spectrum is constant along the wavelength.

However, the above assumption does not hold true for astronomical objects. The structure of the object spectrum may cause a systematic effect in the above procedure because high-frequency components of the spectrum may also shift the position of the source in the reconstructed image. For example, a stellar spectrum is mainly composed of a gradual slope (i.e., blackbody function) and many absorption lines. While the blackbody function does not affect the procedure due to its low-frequency component in the spectral regime, the absorption lines contain high frequency components. In other words, the Fourier transformation of the stellar spectrum is expressed by the sum of the main peak in the center and multiple subpeaks that are entirely spread across the Fourier regime. The true complex amplitude can be calculated if the main peak exceeds the subpeaks. Otherwise, the high-frequency components of the spectrum may shift the position of the source in the reconstructed image, causing a systematic error in the complex visibility measurement. Thus, the spectral imaging method potentially may have a limitation in extracting the true complex visibility.

In Section 5.4, we will simulate measuring the reflected light from the Europa's disk with a stellar space interferometer, using a theoretical stellar spectrum having many absorption lines. We will demonstrate whether or not the structure of the object spectrum affects the true complex visibility measurement.

**4. Requirements for Measurement of the Complex Amplitude**

As discussed in Section 2.1, the DPSI mitigates the requirements for generating the interference fringes. This section analytically describes the requirements for measuring the complex amplitude at each $U$-$V$ point by developing the analytical description of the spectral imaging method in Section 3. The requirements imposed on each exposure measurement and over the entire observation period of an astronomical object are derived in Sections 4.1 and 4.2, respectively. Finally, the requirement imposed on the phase shifter to minimize the systematic error of the complex amplitude measurement is quantitatively evaluated in Section 4.3.

## 4.1. Requirement imposed for each exposure measurement

The measured complex amplitude under the optical path and angular differences between the two beams is described in Equation (9). The true complex amplitude from the astronomical object is, in principle, reconstructed by compensating the systematic effects. However, the optical path and angular differences between the two beams continuously change even during one short exposure time. In this case, Equation (9) is rewritten as

$$\int dt \left\{ I(x,y,\vec{\theta}) - 2|E_0(x,y,\vec{\theta})|^2 + i(I'(x,y,\vec{\theta}) - 2|E_0(x,y,\vec{\theta})|^2) \right\}$$

$$= V(\vec{\theta},\lambda) \int dt \, e^{i\frac{2\pi}{\lambda}((\alpha_2 - \alpha_1(t))x + (\beta_2 - \beta_1(t))y + (c_1 - c_2(t)))}, \quad (12)$$

where the true complex amplitude, $V(\vec{\theta},\lambda)$, is assumed to be constant. The complex amplitude obtained through Equation (12) decreases as the exposure time increases because the interference fringe is blurred due to the continuous change in the optical path and angular differences. Thus, the real and imaginary parts of the complex amplitude for each exposure data can be obtained at more than three sigma levels under the following condition:

$$\frac{Re, Im\left[\int dt \, e^{i\frac{2\pi}{\lambda}((\alpha_2 - \alpha_1(t))x + (\beta_2 - \beta_1(t))y + (c_1 - c_2(t)))}\right]}{\int dt} \geq \frac{3}{\sigma_{S/N,exposure}}, \quad (13)$$

where $Re[C]$ and $Im[C]$ are the real and imaginary parts of the complex value, $C$, and $\sigma_{S/N,exposure}$ is the measurement signal-to-noise ratio of each spectral element taken at an exposure time. Importantly, the integration range corresponds to each exposure time. The signal-to-noise ratio is defined as the ratio of the incident signal from the object to the shot noise originated from the object and detector. When the requirement of Equation (13) is satisfied, the complex amplitude for each spectral element can be measured at more than three sigma levels. The achievable signal-to-noise ratio is larger for bright sources. As a result, the large attenuation of the complex amplitude measurement, which is caused by the large change in the optical path and angular differences during one exposure time, is accepted. The above requirement, which limits each exposure time, is imposed on all beam combiners (including the DPSI).

*4.2. Requirement imposed over the observation period of each astronomical object*

To obtain the three-dimensional structure of the source (i.e., two-dimensional intensity distribution and one-dimensional spectrum), we observe the complex amplitude over various baseline vectors so that the entire *U-V* plane is filled. The DPSI is additionally imposed by the following two conditions related to the acceptable optical path and angular differences over the observation period of an astronomical object. Regarding the acceptable optical path difference between the two beams, DPSI must keep its coherent state over the observation period. When the optical path difference between the two beams is $c_1 - c_2$, the degradation of the complex visibility, $\delta V$, is written as

$$\delta V(c_1 - c_2) = 1 - \left|\mathrm{sinc}\left(\frac{\pi(c_1-c_2)}{R\lambda}\right)\right|. \tag{14}$$

where $R$ is the resolving power of DPSI. To measure the complex amplitude for each spectral element at more than three sigma levels, the acceptable optical path difference must be satisfied with the following Equation:

$$\left|\mathrm{sinc}\left(\frac{\pi(c_1-c_2)_{max}}{R\lambda}\right)\right| \geq \frac{3}{\sigma_{S/N,measurement}}, \tag{15}$$

where $(c_1 - c_2)_{max}$ is the acceptable optical path difference for measuring the complex amplitude at more than three sigma levels, and $\sigma_{S/N,coadd}$ is the signal-to-noise ratio of each spectral element after coadding the measurements at each *U-V* point.

In the conventional beam combiner, the relative angle over the observational period is acceptable if the point-spread functions of the two beams overlap on the focal plane. In other words, the relative tilt between the two beams should be less than the angular resolution of each beam. In contrast, because the detector plane of DPSI is optically conjugated to the interference plane (i.e., the beam splitter), the interference fringe can be formed once the two parallel beams are overlapped on the beam splitter. When the distance between the positions of the two parallel beams on the interference plane is less than 0.1 times the beam diameter, the interference efficiency exceeds 0.9. In this case, the relative angles between the two beams around the two axes are given by

$$\alpha_2 - \alpha_1, \beta_2 - \beta_1 \leq \frac{1}{5}\frac{D}{|\vec{B}|}. \tag{16}$$

When $D = 5\ cm$ and $|\vec{B}| = 50\ m$, the relative angle is 40 arcseconds, which is much larger than the acceptable one for the conventional beam combiner. However, if each spectral element is not fully sampled by the detector, the acceptable relative angle of the DPSI is determined so that the inference fringe of each pixel is not blurred by the large angular difference between the two beams. Because the relative angle continuously changes within a pixel due to the finite number of samplings for each spectral element, the complex amplitude recorded in each pixel is affected by the relative angle between the two beams. The proper relative angle is derived by the following equation:

$$\left|\frac{\iint dxdy\, e^{i\frac{2\pi}{\lambda}((\alpha_2-\alpha_1(t))x+(\beta_2-\beta_1(t))y)}}{\iint dxdy}\right| \geq \frac{3}{\sigma_{S/N,measurement}}, \tag{17}$$

where the integration range is the size of each pixel. Given that each spectral element is sampled by the number of $n_{pix}$ along one direction, the pixel size corresponds to $\frac{D}{n_{pix}}$ on the incident aperture plane, and Equation (17) is written as

$$\left|sinc\left\{\frac{\pi}{\lambda}\frac{D}{n_{pix}}(\alpha_2-\alpha_1)_{max}\right\}sinc\left\{\frac{\pi}{\lambda}\frac{D}{n_{pix}}(\beta_2-\beta_1)_{max}\right\}\right| \geq \frac{3}{\sigma_{S/N,measurement}}, \tag{18}$$

where $(\alpha_2 - \alpha_1)_{max}$ and $(\beta_2 - \beta_1)_{max}$ are the acceptable angular differences around the $y$ and $x$ axes over the observation period of one astronomical object in the unit of $\frac{\lambda}{D}$. Thus, the acceptable ranges of the optical path and angular differences are determined from the requirements described by Equations (15), (16) and (18), respectively. The above requirements of a DPSI mounted on SEIRIOS are quantitatively estimated as an example in Section 5.3.

*4.3. Requirement on the phase shifter*

By adding an achromatic phase of π/2, we can measure the complex amplitude from one exposure measurement. However, the achromatic phase cannot be realized over a wide wavelength range. Here, we analytically derive the impact of the phase-shift error $\Delta\phi$ on the complex amplitude measurement. When $\Delta\phi$ is small, Equation (9) is rewritten as

$$I(x, y, \vec{\theta}) - 2|E_0(x, y, \vec{\theta})|^2 + i(I'(x, y, \vec{\theta}) - 2|E_0(x, y, \vec{\theta})|^2)$$
$$\approx V(\vec{\theta}, \lambda) + i\Delta\phi(V(\vec{\theta}, \lambda) + c.c.), \tag{9}$$

where we have assumed no optical path difference and zero relative angle for simplicity. The the phase shift error introduces a systematic error in the measured complex visibility. As the systematic error is approximately proportional to the phase-shift error, we assume that we can accept errors up to 10% of the original phase shift (i.e., π/2).

## 5. Application to a formation flying interferometer using small satellites

*5.1 Overview of SEIRIOS*

We briefly introduce the concept of a formation-flying stellar interferometer composed of one 50 *kg*-class microsatellite and two 6U CubeSats for achieving high spatial resolution spectral imaging. In this article, the 50 *kg*-class microsatellite and 6U CubeSats are named "mother" and "daughter" satellites, respectively. Figure 5 summarizes the concept of SEIRIOS, which will be launched in the middle of the 2020s. An interferometric fringe from an astronomical object with a long baseline constructed by multiple apertures will be detected on-orbit for the first time. The main purpose of SEIRIOS is to construct a formation-flying interferometry using multiple small satellites in a low Earth orbit. The three satellites are combined at the launch, and they are separated on-orbit after performing the calibration of the measurement system so that an interference fringe of two beams from the daughter satellites is acquired under the docked state. SEIRIOS is operated in a Sun-Synchronous Orbit (SSO) to be thermally stabilized. Considering that some of the objects in the Solar system and very few nearby bright galactic objects (i.e., stars) are almost aligned through the connecting line between the Sun and SEIRIOS, the Solar system objects and the galactic objects on the opposite side of the Sun are suitable targets of SEIRIOS. The interferometric fringes from the objects can be continuously measured for a certain period. The *U-V* plane is covered by changing the baseline length, $B$, from 10 to 100 $m$, corresponding to imaging spatial resolutions (i.e., $1.22 \frac{\lambda}{B}$) of 20 to 2 milli-arcseconds at a wavelength, $\lambda = 1\ \mu m$.

SEIRIOS applies DPSI as the beam combiner and continuously obtains a wide wavelength range (0.6 to 1.7 $\mu m$) in one exposure measurement. The near-infrared wavelength range includes two strong water vapor absorption bands in the ranges of $0.99 - 1.09\ \mu m$ and $1.35 - 1.51\ \mu m$, which cannot be observed from ground-based telescopes. The resolving power is approximately 1000 over the entire wavelength range except for the edge of each order, and the strong water vapor absorption lines included in the wavelength range can be resolved. If the target object emits the water vapor to outer space, such as Europa[34,35,36] and Enceladus[37], high-spatial resolution spectral imaging may identify the place where the water vapor is generated.

We employ an optimized spectral imaging method for the DPSI. As described in Section 3, the method allows the extraction of the true complex visibility from a spectrally resolved interference fringe taken at an exposure time. The spectral imaging method provides two main advantages in a SEIRIOS application. First, the true complex visibility can be extracted from one exposure measurement, so the spectral imaging method is fully functional even under unstable observing conditions leading to rapid changes in the optical path difference and relative angle between the two beams. Second, the signal-to-noise ratio of the true complex visibility at each *U-V* point is imposed by coadding the complex visibilities in short exposure measurements. We note that Ikari *et al.* will submit an alternative paper on the satellite bus of SEIRIOS from a space engineering perspective.

*5.2 Optical design*

Figure 6 shows the optical design of SEIRIOS under the docked state. The fore-optics and the densified pupil spectrograph for recording the interference fringes are placed in the upper and lower layers, respectively. The light coming from an astronomical object is first reflected by a 45-*degree* rectangular plane mirror mounted on each daughter satellite, and each beam is introduced to the mother satellite. One side of the square beam is 50 *mm* at the entrance of the mother satellite and is reduced to 10 *mm* before the two beams interfere. The two beams interfere on the plane of the polarizing beam splitter in the mother satellite after an achromatic phase shifter is inserted into one of the two beams. As explained in Section 3, the achromatic phase shifter is composed of two linear polarization plates and a quarter-wave plate sandwiched between them. Because a different phase shift is given to each polarization state, the polarizing beam splitter extracts one of the two polarization states. While one of the two interfered beams is introduced to a spectrograph for recording the spectrally resolved interference fringes, another is introduced to two sensors (an attitude sensor for measuring the relative tilt between two daughter satellites and a position sensor for measuring the relative position between the two beams on the interfered plane) after the interfered beam is divided into two with another beam splitter. The two sensors work at a wavelength ranging from 700 to 1000 *nm*. After passing through the beam splitter, one of the interfered beams is focused on the attitude sensor with a camera lens system. The pixel scale of the attitude sensor is 0.5"; hence, the attitudes of the satellites can be measured at a sub-arcsecond level provided that the sensor fully collects the photons. The effective field of view of the sensor is limited by vignetting owing to the clear apertures of the daughter and mother satellites, depending on the baseline. The effective field-of-view depends on the baseline length and is approximately 3' in diameter when the baseline length is 15 *m*. Moreover, two square pupil images with a one-side length of 5 *mm* are formed on the position sensor, and each image is sampled with approximately 1500 *pixels*. Therefore, the relative position can be estimated with an accuracy of less than 0.01 *mm*. Based on the information obtained by the two sensors, the relative tilt and position between the two beams are compensated by the attitude/position control systems of SEIRIOS. The relative attitude and position of the two beams on the interference plane are roughly compensated by controlling the daughter satellites and precisely by controlling an internal plane mirror mounted on one of the two daughter satellites.

One of the two interfered beams is introduced to the densified pupil spectrograph to record the spectrally resolved interference fringes. A field stop on the focal plane in the optical relay system determines the field of view of the spectrograph to 1' in diameter. Next, the beam is divided into five with a compact beam slicer, named Bowen-Walraven type image slicer[27,28]. After the divided beams are collimated with a Cassegrain optical system composed of three mirrors, the beam is divided into two with a dichroic mirror in terms of the wavelength: 0.6 – 1.0 *μm* for the short channel and 1.0 – 1.7 *μm* for the long channel. While the first-order spectrum is obtained on the detector of the short channel with a transmission grating, multiple-order spectra are obtained by passing through an Echelle spectrograph composed of a transmission grating and a cross-disperser that separates the spatially overlapped multiple orders. The resolving powers for the short and long channels are approximately 1000 and 1100 at the center of each order, respectively. Importantly, the wavelength range of the short channel is adjustable by moving its detector along the dispersion direction. The wavelength range is selected such that the target image is reconstructed through performing the Fourier transformation of the complex visibility taken at one exposure data (see Sections 2.2 and 3.2). Because the wavelength range of the short channel is limited, the reconstructed image is sometimes nulled by a combination of the baseline length, the diameter of the target, and the wavelength range.

An electron-multiplying charge-coupled device (EMCCD) and an indium gallium arsenide (InGaAs) detector are applied as the science detectors for the short and long channels, respectively. The detectors for the short and long channels are assumed to be iXon888 and PIRT1280A1-12 provided by the Oxford Instrument and the Princeton Infrared Technologies, Inc., respectively. The detector formats for the short and long channels are 1024 x 1024 and 1280 x 1024 pixels with pixel pitches of 13 and 12 $\mu m$, respectively. When the EMCCD and InGaAs detectors are cooled down to 220 $K$ by a combination of passive cooling and Peltier elements, the dark current and read noise of the InGaAs detector are 160 $e$-/sec and 35 $e$-, respectively. Both the detector noises of EMCCD are negligible (less than 1 $e$-) because of a high EM gain. As will be discussed, the low detector noise of the short channel allows us to measure the optical path difference between the two beams even from each short exposure data. The total throughput of DPSI, including 50 % throughput of the beam splitter, was roughly estimated to be 0.3, assuming that the reflectivity of each mirror is 0.985, the diffraction efficiency is 0.8, the throughput of each lens is 0.99, and the quantum efficiency is 0.85. Figure 7 shows the Echelle formats of the two science detectors. The sizes of a divided slice image for the short and long channels are approximately 10 x 2 and 18.5 x 3.7 pixels on the detectors, respectively. The smaller number corresponds to that of the samplings along the wavelength. Because the number of slices for each channel is five, the numbers of each spectral element for the short and long channels are set to approximately 100 and 320 pixels, respectively. Thus, SEIRIOS can continuously cover a wide wavelength range from 0.6 to 1.7 $\mu m$ in single exposure time if the phase shifter meets the requirement introduced in Section 4.3. We should note, however, that the achromatic phase shifter may limit the wavelength range because whether the phase of $\pi/2$ is practically realized over a wide wavelength range at the same time is still unknown. The basic parameters of SEIRIOS are listed in Table 1.

## 5.3. Requirements of formation-flying interferometry

We quantitatively evaluate the acceptable relative angle and optical path difference between the two beams of DPSI in the SEIRIOS application. The requirements imposed over the entire observation period of an astronomical object are analytically described in Equations (15), (16), and (18) in Section 4. Figure 8 shows the acceptable optical path difference and angular difference between the two beams at 750 and 1300 $nm$ over the observation period of an astronomical object for SEIRIOS with a baseline length of 20 $m$. When the signal-to-noise ratio of each spectral element after coadding the measurements at each $U$-$V$ point, $\sigma_{S/N,coadd}$, reached 10, the acceptable optical path difference and angular difference for reconstructing the complex visibility at the three sigma levels are $\pm 700$ $\mu m$ and $\pm 20"$, respectively, for the short channel and $\pm 1200$ $\mu m$ and $\pm 50"$, respectively, for the long channel. The acceptable angular difference is almost ten times larger than the diffraction limit of each aperture, corresponding to 3" at 750 $nm$ and 5".2 at 1300 $nm$. This is because the spectra formed on the detectors of the short and long channels are sampled by 10 and 18 pixels along one direction, respectively. Note that the acceptable relative angle in the long channel is limited by the degradation of the interference efficiency when the two beams are displaced on the interference plane (see Equation (16)).

The requirement imposed on each exposure measurement is quantitatively evaluated by Equation (13). Figure 9 shows the acceptable amount of change in the optical path difference during one exposure time. Here, the optical path difference is assumed linear with respect to time. When the optical path difference changes by one-half the observing wavelength, the interference fringe is almost smoothed out and cannot be obtained from one exposure measurement. In contrast, the degradation of the interference fringe measurement is fully suppressed when the optical path difference changes by one-third of the observing wavelength. Thus, one-third of the observing wavelength, corresponding to approximately 0.25 and 0.45 µm in the short and long channels, respectively, is the acceptable optical path difference. Based on these considerations, the requirements for measuring the complex amplitude are summarized in Table 2.

*5.4. Simulation*

Finally, we investigate the applicability of the proposed spectral imaging method can be applied to the Europa's disk observations under a simple assumption, namely, that the length of each exposure time is limited to the order of 10 milliseconds. To extract the true complex visibility from the interference fringe taken at each exposure time, we require a clear object image constructed by performing the Fourier transformation of the complex visibility along the wavelength. The object's position on the reconstructed image represents the optical path difference between the two beams. As discussed in Section 3.3, the structure of the object spectrum may affect the true complex visibility measurement. We investigate whether or not the proposed spectral imaging method can be validated for measuring the solar system objects using a realistic stellar spectrum model. Moreover, we investigate whether the signal-to-noise ratio for each exposure data is sufficient to reconstruct the astronomical image or not by performing numerical simulations. Since the main purpose of the SEIRIOS mission is to search for water vapor on the Solar system objects using the strong water absorption bands at the near-infrared wavelengths, Jupiter's icy moon Europa is taken as the target for the feasibility study.

The parameters for the simulations are set as follows: The baseline length is fixed to 20 *m*. Considering that the detector for the long channel is much higher than that for the shorter channel, the wavelength range for measuring the optical path difference is set to 0.7 - 1.0 *μm*. The dark current and read noise are set to 0.1 *e-/sec* and 0.04 *e-*, respectively. The latter would be achievable when a high EM gain (~ 1000) of the EMCCD is applied to the shorter channel. We note that the read noise of iXon Ultra 888 is 40 *e-* without the EM gain for 10 MHz readout. The resolving power is set to 1000 over the wavelength range. The target object is simply positioned to the center of the field of view. In other words, the vector between the target object and the interferometer is perpendicular to the baseline. Figure 10 (a) shows Europa's disk in the sky. The angular diameter of Europa's disk is approximately 1".0, and the disk is shined by the reflected light from the Sun. The surface albedo is simply set to be constant of 0.8 over the wavelength range; the spectrum of the Europa disk is the same as that of the Sun. A high spectral resolution spectrum of the Sun was generated by the *BT-settl model*[41] with the surface gravity of 4.5 log (*cm/s²*), effective temperature of 5800 *K*, and solar metallicity and was convolved with a Gaussian line spread function[42] of $exp\left[-(4ln2)R^2\left(\frac{\lambda}{\lambda_0}-1\right)^2\right]$, where $\lambda_0$ is the central wavelength of each spectral element.

We first generate the spectrally resolved data without any noises. Figure 10 (b) shows the reconstructed image obtained by performing the Fourier transformation of the data along the wavelength. The diameter of the object in the reconstructed image corresponds to the angular diameter of the input object. The object's offset is 1".0, which is equal to the ratio of the optical path difference to the baseline length. We confirmed that the offset is not affected by the solar spectrum, which has many absorption lines and a gradual slope, because the low-frequency component included in the object spectrum surpasses the other high frequency components. Thus, the optical path difference between the two beams can be principally measured by this method. We next included the shot noise from the object and the detector noise in the spectrally resolved data. Figure 10 (c) shows the reconstructed image from the simulated data. The reconstructed Europa disk is affected by the shot noise and detector noise. To improve the signal-to-noise ratio, we apply the moving average to the reconstructed image. The center position of the Europa disk can be detected in the reconstructed image applying the moving average (see Figure 10 (d)). Thus, we applied the method to the Europa observation and successfully measured the optical path difference from one short exposure data. After the optical path difference between the two beams is calibrated for each short-exposure data, the compensated complex visibility must be coadded to achieve a good signal-to-noise ratio for each spectral element at each *U-V* point.

## 6. Conclusion

We proposed a new type of beam combiner for space interferometers named the "DPSI," a development of the densified pupil spectrograph interferometer. In the DPSI, the interfered beam formed on the interference plane is transferred to the detector. With this beam combiner, we can reconstruct the spectrally-resolved complex visibility from one exposure measurement without delay-line scanning and chopping interferometry. In addition, because the detector plane is optically conjugated to the interference plane, the beam combiner can directly measure the relative phase difference between the two beams, thereby mitigating the requirements of measuring complex visibility. In other words, the interference fringes can be obtained even if the angular difference between the two beams is much larger than the diffraction limit of each aperture. A larger optical path difference between the two beams is also acceptable because the spectrograph has moderate resolving power. We then developed a new spectral imaging method and optimized it for the new beam combiner. The true complex amplitude could be derived from one exposure measurement by compensating the relative angular and optical path differences between the two beams only when the following two conditions were satisfied: (1) an astronomical object within the stellar interferometer's field of view is obtained with a modest signal-to-noise ratio, and (2) the structure of the object spectrum does not affect the true complex visibility measurement. Thus, the proposed spectral imaging method will be limited by the observing conditions and astronomical objects.

We applied the new beam combiner and the spectral imaging method to a free-flying space interferometer called SEIRIOS, which is composed of one 50-*kg* class satellite and two 10-*kg* class CubeSats. The main purpose of the SEIRIOS mission is to validate the formation flying interferometry in observations of astronomical objects, including some of the Solar System objects and nearby bright galactic objects. The SEIRIOS mission covers a wide wavelength range (0.6 - 1.7 μm) at one time with two spectral channels: a short channel (0.6 – 1.0 μm) and a long channel (1.0 – 1.7 μm). This wavelength range includes strong water absorption bands at near-infrared wavelengths. The SEIRIOS mission system comprises two measurement systems: one for measuring the angular difference and relative position between two beams on the interference plane and DPSI for recording the spectrally resolved interference fringes. Based on the specification of the spectrograph, the acceptable optical path and angular differences are $\pm 700$ μ$m$ and $\pm 20"$ for the short channel and $\pm 1200$ μ$m$ and $\pm 50"$ for the long channel. Thus, DPSI can mitigate the requirements on the relative position and angular difference among the three spacecraft.

Finally, we investigated the efficacy of the proposed spectral imaging method in Europa observation, assuming that each exposure time is limited to the order of 10 milliseconds. We found that Europa's disk was successfully reconstructed over a short-exposure time of 10 milliseconds. The reflected spectrum of the Europa's disk (i.e., the Sun's spectrum) did not affect the measurement of the true complex visibility because the low frequency component in the spectrum exceeds the high frequency ones. The optical path difference could be estimated by measuring the shift in the Europa's disk on the reconstructed image. The object was shifted on the reconstructed image by the ratio of the optical path difference to the baseline length along the baseline direction. Therefore, SEIRIOS can extract the true complex visibility without impacting the optical path difference and angular difference from each exposure data. The SEIRIOS mission may be able to perform high-resolution spectral imaging of the Europa's disk and bright astronomical objects after filling the *U-V* plane by changing the baseline length. As the next step, we investigate what types of astronomical objects the proposed method can be applied to. We also numerically simulate the formation-flying interferometry and hence evaluate its detailed performance in astronomical SEIRIOS observations.

*Acknowledgments*

We sincerely appreciate two anonymous referees for providing useful and valuable comments that largely enhanced the value of this manuscript. T.M. was supported by Grant-in-Aid from MEXT of Japan, No. 19H00700.

**Taro Matsuo** is an associate professor of physics at Nagoya University. His research focuses on the field of exoplanet science, with an emphasis on developing several instrumental concepts for the characterization of nearby habitable planet candidates, including transit spectrograph, interferometer, and coronagraph.

**Satoshi Ikari** is an assistant professor in the Department of Aeronautics and Astronautics at The University of Tokyo. His research focuses on astrodynamics and the attitudes and orbit control of satellites. In particular, he studies the application of formation flying satellites and the development of microsatellites.

**Hirotaka Kondo** is a graduate student at the University of Tokyo. He studies spacecraft formation flying, particularly the accurate control of the relative position and attitude of spacecraft for space interferometers, including ground facilities for validation of the control methods.

**Sho Ishiwata** was a former graduate student at the University of Tokyo. His research area include image reconstruction methods for space interferometers by Earth-orbiting formation flying and the conceptual design of SEIRIOS.

**Shinichi Nakasuka** is a professor of Department of Aeronautics and Astronautics at University of Tokyo. His research interests include micro/nano/pico-satellite systems and utilizations and novel space systems. Especially he already succeeded in launching thirteen satellites less than 100kg including the world first 1U CubeSat.

**Tomoyasu Yamamuro** is an optical engineer. He designs optical and infrared astronomical optical system, including multiband imager, high resolution spectrographs, and pupil slicer systems.

**Table 1** Configuration of SEIRIOS.

| Item | Parameter |
| --- | --- |
| Configuration | 50 *kg*-class microsatellite and two 6U CubeSats |
| Life span | 0.5 – 1 year |
| Baseline length | 10 – 100 *m* |
| Collecting area of each aperture | 25 *cm²* (a rectangular aperture with a one-side length of 5 *cm*) |
| Observation wavelength | 0.6 – 1.0$^{(*)}$ *μm* (optical), 1.0 – 1.7 *μm* (near-infrared) |
| Resolving power | 1100 (optical), 1000 (near-infrared) at the center of each order |
| Field of view of the spectrograph | 1' in diameter |
| Total throughput | 0.3 |
| Type of spectrograph | Densified pupil spectrograph |
| Type of monitor | Attitude sensor and position sensor |
| Type of detector | EMCCD (optical), InGaAs (near-infrared) |
| Detector format | 1024 x 1024 (optical), 1280 x 1024 (near-infrared) |
| Pixel pitch | 13 *μm* (optical), 12 *μm* (near-infrared) |
| Number of samplings | 92 pixels (optical), 320 pixels (near-infrared) |
| Dark current | < 0.1 *e-/sec* (optical), 160 *e-/sec* (near-infrared) |
| Read noise | < 1 *e-/sec* (optical), 35 *e-* (near-infrared) |

(*) The range of the spectrum obtained at one measurement is limited to approximately 0.3 *μm*. The wavelength of the short channel is changeable by shifting the optical detector along the dispersion direction.

**Table 2** Requirements on the optical path and angular differences between the two beams for SEIRIOS.

| Item | Parameter |
| --- | --- |
| 1. Acceptable optical path difference over the observation period | 700 µm at 750 $nm$, 1200 µm at 1350 $nm$ (Equation (14)) |
| 2. Acceptable angular differences for not smearing the interference fringe around two axes, $\alpha_2 - \alpha_1$ and $\beta_2 - \beta_1$ | $\pm 20"$ at 750 $nm$, $\pm 40"$ at 1350 $nm$ (Equation (18)) |
| 3. Acceptable angular differences for overlapping two beams on the interference plane around two axes, $\alpha_2 - \alpha_1$ and $\beta_2 - \beta_1$ | $\pm 50"$ and $\pm 20"$ at the baselines of 20 and 50 *m* (Equation (16)) |
| 4. Acceptable optical path and angular differences for each exposure | 0.25 µm at 750 $nm$ and 0.45 µm at 1350 $nm$ (Equation (13)) |

**Caption List**

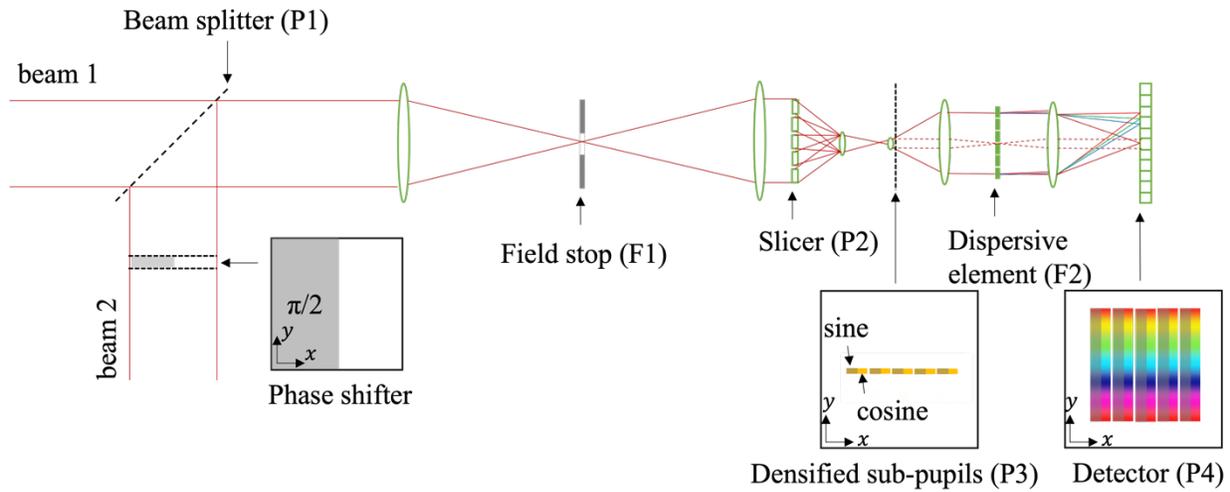

**Figure 1.** Conceptual diagram of the densified pupil spectrograph interferometer (DPSI). P and F represent the pupil and focal planes, respectively. The beam splitter (i.e., interference plane of two beams) is optically conjugated to the pupil slicer, densified sub-pupils, and detector plane. The phase shift of $\pi/2$ indicated by the gray shaded region is added to half of the entrance pupil for the beam 2 before the two beams are interfered. The sine component of the complex visibility is obtained in half of the interfered beam. We assume that the densified pupil spectra are aligned along the y axis.

(1) Data acquisition (Section 3.1)

Measurement of the cosine and sine components for each pixel and each spectral element, $(x_j, y_k, \lambda_l)$

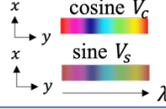

$V_c(x_j, y_k, \lambda_l), \ldots, V_c(x_{n_{pix}}, y_{n_{pix}}, \lambda_n)$

$V_s(x_j, y_k, \lambda_l), \ldots, V_s(x_{n_{pix}}, y_{n_{pix}}, \lambda_n)$

(2) Reconstruction of complex visibility (Section 3.2)

Reconstruction of the complex visibility for each pixel and each spectral element, $(x_j, y_k, \lambda_l)$ from the cosine and sine components:

$V(x_j, y_k, \lambda_l) = V_c(x_j, y_k, \lambda_l) + iV_s(x_j, y_k, \lambda_l), \ldots, V(x_{n_{pix}}, y_{n_{pix}}, \lambda_l) = V_c(x_{n_{pix}}, y_{n_{pix}}, \lambda_l) + iV_s(x_{n_{pix}}, y_{n_{pix}}, \lambda_l)$

$\vdots$

$V(x_j, y_k, \lambda_n) = V_c(x_j, y_k, \lambda_n) + iV_s(x_j, y_k, \lambda_n), \ldots, V(x_{n_{pix}}, y_{n_{pix}}, \lambda_k) = V_c(x_{n_{pix}}, y_{n_{pix}}, \lambda_n) + iV_s(x_{n_{pix}}, y_{n_{pix}}, \lambda_n)$

(3) Measurement of relative tilt angles (Section 3.3)

Measurement of the relative tilts between two beams from the phase distribution of the complex visibility for each spectral element:

$\arg\{V(x_j, y_k, \lambda_l)\}, \ldots, \arg\{V(x_{n_{pix}}, y_{n_{pix}}, \lambda_l)\}$

$\vdots$

$\arg\{V(x_j, y_k, \lambda_n)\}, \ldots, \arg\{V(x_{n_{pix}}, y_{n_{pix}}, \lambda_n)\}$

$\rightarrow \alpha_2 - \alpha_1, \beta_2 - \beta_1$

(4) Calibration of complex visibility in terms of relative tilt angles (Section 3.3)

Coadding the complex visibility for each spectral element after compensating the relative tilt angles for each pixel:

$$V_{coadd}(\lambda_l) = \sum_{j,k}^{n_{pix}} V(\lambda_l) e^{-\frac{2\pi i}{\lambda_l}\{(\alpha_2-\alpha_1)x_j + (\beta_2-\beta_1)y_k\}}, \ldots, V_{coadd}(\lambda_n) = \sum_{j,k}^{n_{pix}} V(\lambda_n) e^{-\frac{2\pi i}{\lambda_n}\{(\alpha_2-\alpha_1)x_j + (\beta_2-\beta_1)y_k\}}$$

(5) Measurement and calibration of optical path difference (Section 3.3)

Measurement of the optical path difference between two beams, $c_2 - c_1$, through performing Fourier transformation of the coadded complex visibility along the spectral direction:

$FT\{V_{coadd}(\lambda_l)\}_{spectrum} \rightarrow c_2 - c_1$

Derivation of the true complex visibility by calibrating the optical path difference between two beams:

$$V_{true}(\lambda_l) = V_{coadd}(\lambda_l) e^{-\frac{2\pi i}{\lambda_l}(c_2-c_1)}, \ldots, V_{true}(\lambda_n) = V_{coadd}(\lambda_n) e^{-\frac{2\pi i}{\lambda_n}(c_2-c_1)}$$

**Figure 2.** Procedure for derivation of the true complex visibility for each spectral element. We assume that the pupil for each spectral element is sampled by $n_{pix} \times n_{pix}$ pixels. The number of the spectral elements is set to $n$. $(x, y)$ is the coordinate system of the pupil plane, and $\lambda$ is the observing wavelength. $\alpha_i$ and $\beta_i$ represent the tilt angles around the $y$ and $x$ axes for the beam $i$, respectively. The arg $\{C\}$ shows the argument of the complex amplitude, $C$.

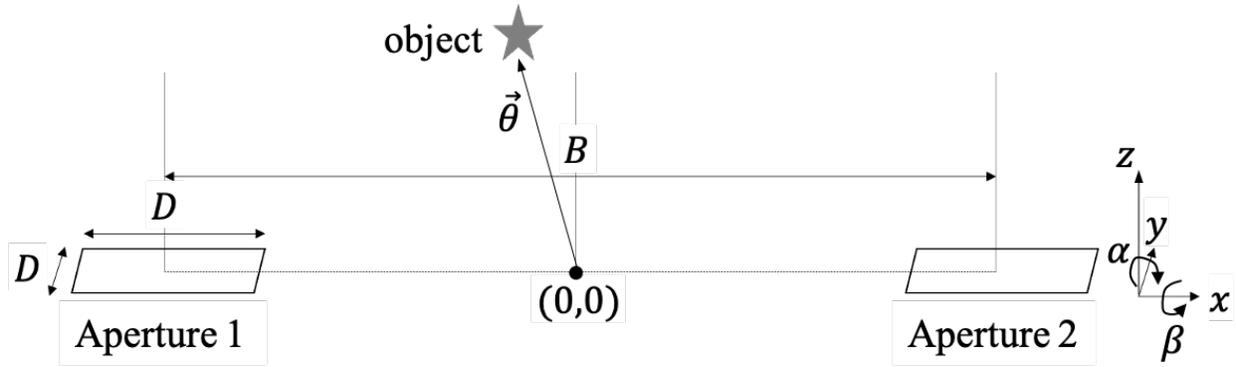

**Figure 3.** Coordinate system and parameters of each entrance aperture.

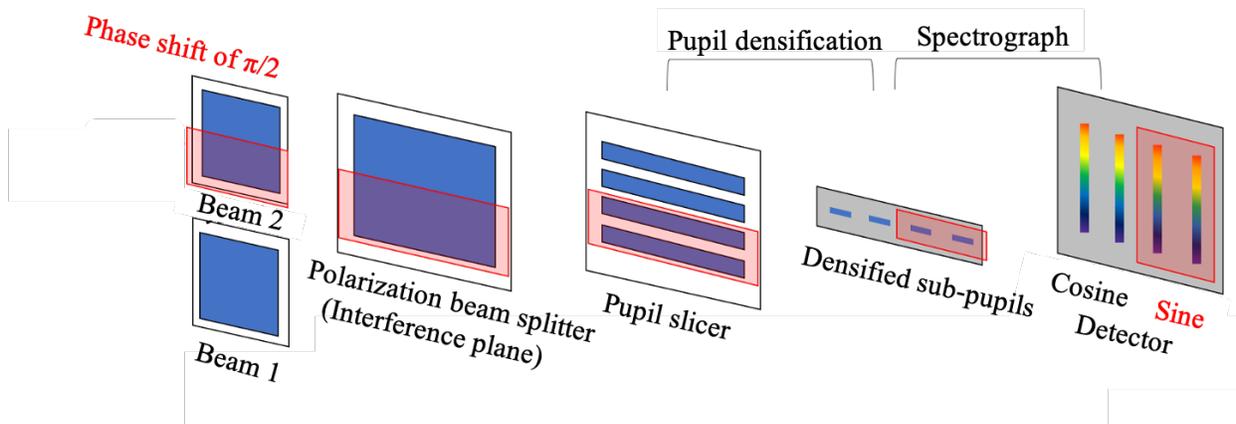

**Figure 4.** Configuration of the achromatic phase shifter optimized for reconstructing the complex amplitude from one exposure data. The blue rectangles represent the beams. A phase shift of π/2 is added to the red area. The cosine and sine components of the complex amplitude are obtained using the science detector.

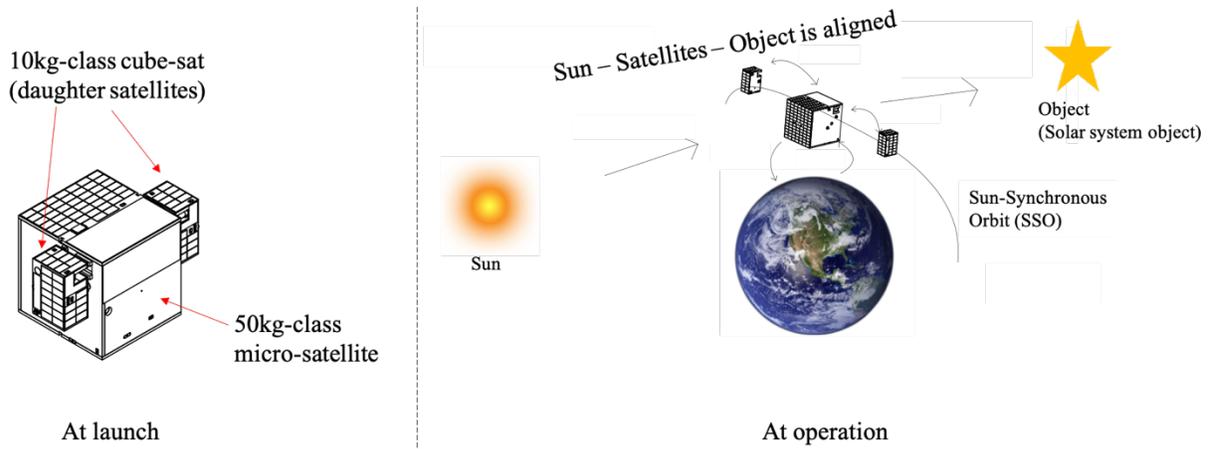

**Figure 5.** Conceptual diagram of SEIRIOS. The left and right panels depict the launch and operation conditions of SEIRIOS, respectively. While two 10 *kg*-class cubesats and one 50 *kg*-class microsatellite are combined at launch, the three satellites are separated. They are formation-flying after the interferometric fringes are obtained in an SSO.

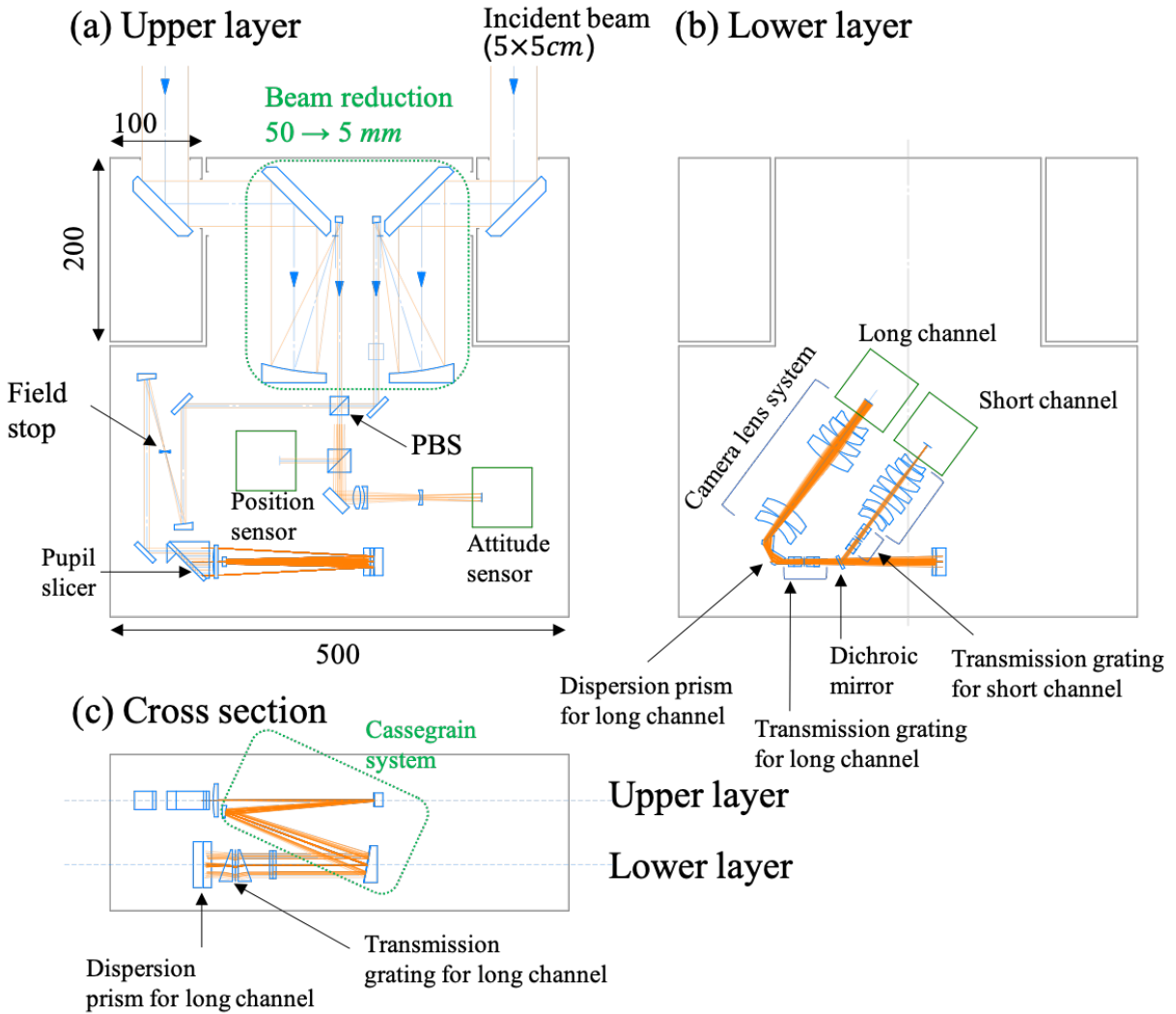

**Figure 6.** Optical design of SEIRIOS under a docked condition. Panels (a) and (b) show the upper and lower layers of the mission equipment, respectively. Panel (c) shows the cross section of the densified pupil spectrograph from the pupil slicer to the dispersion element. The orange lines represent the two beams with a one-side length of 50 *mm*, and the blue arrow shows the flow of the beams. The one-side length of each beam coming from an astronomical object is reduced to 5 *mm* with two concave mirrors before the two beams interfere on the plane of the polarizing beam splitter. An achromatic phase shifter is inserted in one of the two beams. One of the two interfered

beams is introduced to the attitude and position (i.e., pupil monitor) sensors. Another is introduced to the densified pupil spectrograph shown in the lower layer.

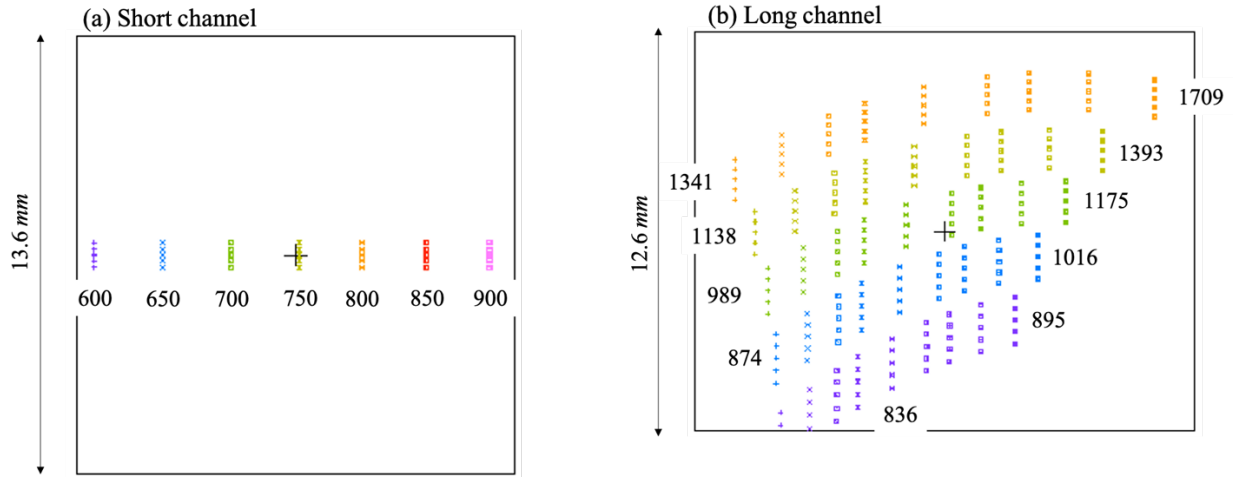

**Figure 7.** Footprints of the (a) short and (b) long channels. The numbers below the spots in Panel (a) show their wavelengths. The numbers shown at each order's left and right edges in Panel (b) represent each order's longest and shortest wavelengths in the *nm* unit. The purple, blue, green, yellow, and orange spectra correspond to 9, 8, 7, 6, and $5^{th}$ orders in Panel (b). The first-order spectrum is formed on the detector for the short channel. The wavelength range of the short channel is adjustable by slightly shifting its detector along the spectral direction. The wavelength could be extended to 1.0 *μm*, and the long channel mainly covers a wavelength range that is longer than 1.0 *μm*.

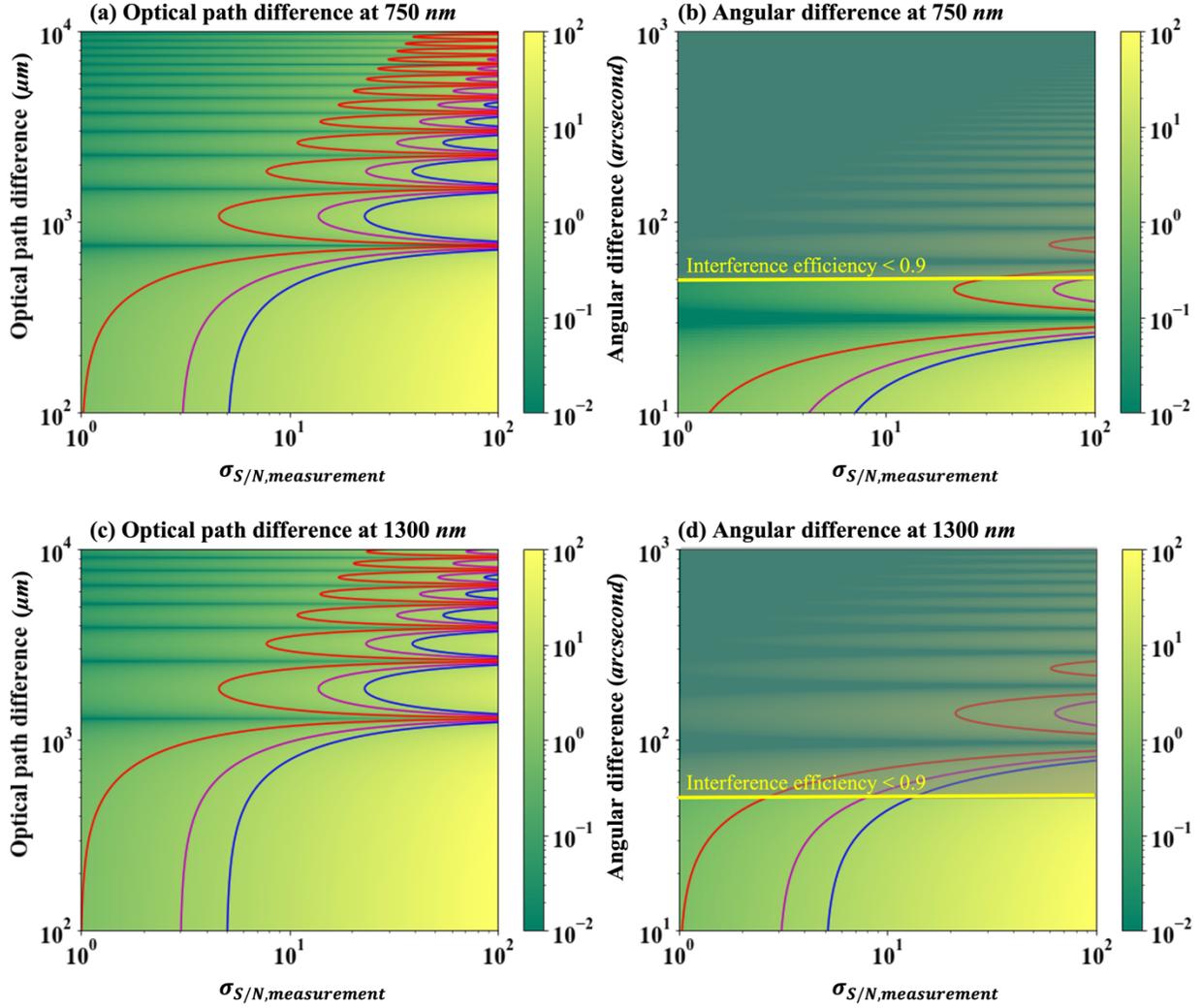

**Figure 8.** (a), (c) Acceptable optical path difference and (b), (d) acceptable angular difference between the two beams for reconstructing the complex amplitude of each spectral element at 750 and 1300 *nm* as a function of the signal-to-noise ratio of each spectral element after coadding the measurements at each *U-V* point, $\sigma_{S/N,coadd}$. When the baseline length was set to 20 *m*, the angular difference should be less than 50" to fully overlap the two beams on the interference plane, and the interference efficiency is 0.9 indicated by yellow horizontal line. The red, magenta, and blue solid lines represent the achievable signal-to-noise ratios of 1, 3, and 5 for reconstructing the complex visibility, respectively. The optical parameters, such as the diameter of the entrance pupil, the spectral resolving power, and the number of the samplings for each spectral element, were set to those of SEIRIOS compiled in Table 2. The wavelengths of 750 and 1300 *nm* represent the short and long channels of SEIRIOS, respectively.

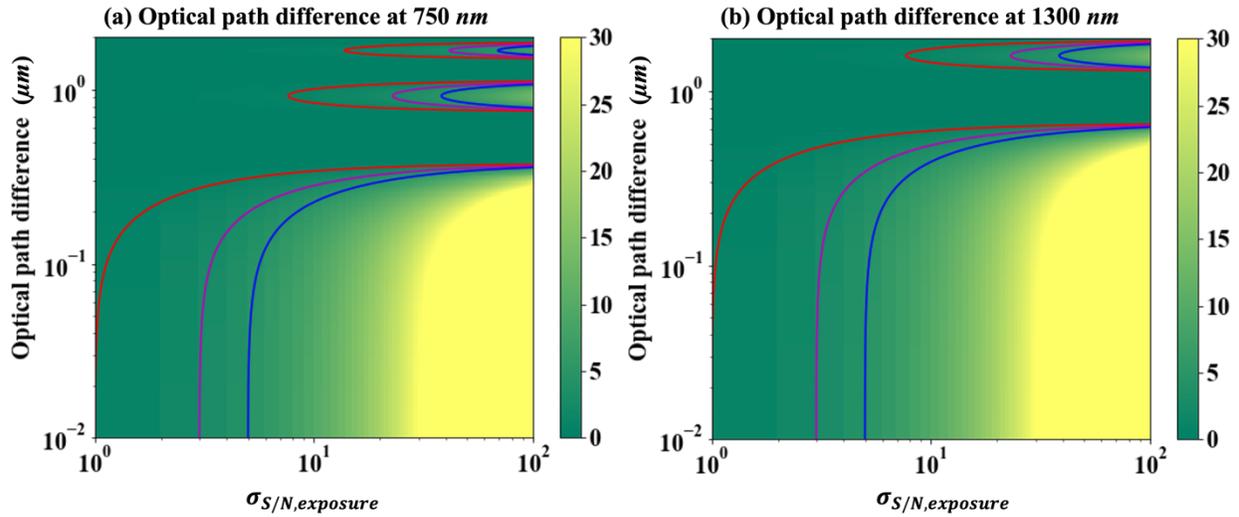

**Figure 9.** Acceptable amount of change in the optical path difference during one exposure time for measuring the real part of the complex visibility at (a) 750 and (b) 1300 *nm* as a function of the signal-to-noise ratio of each spectral element taken at one exposure time. The contour of each panel shows the signal-to-noise ratio for measuring the real part of the complex visibility. The red, magenta, and blue lines represent achievable signal-to-noise ratios of 1, 3, and 5 for the real part of the complex visibility measurement, respectively.

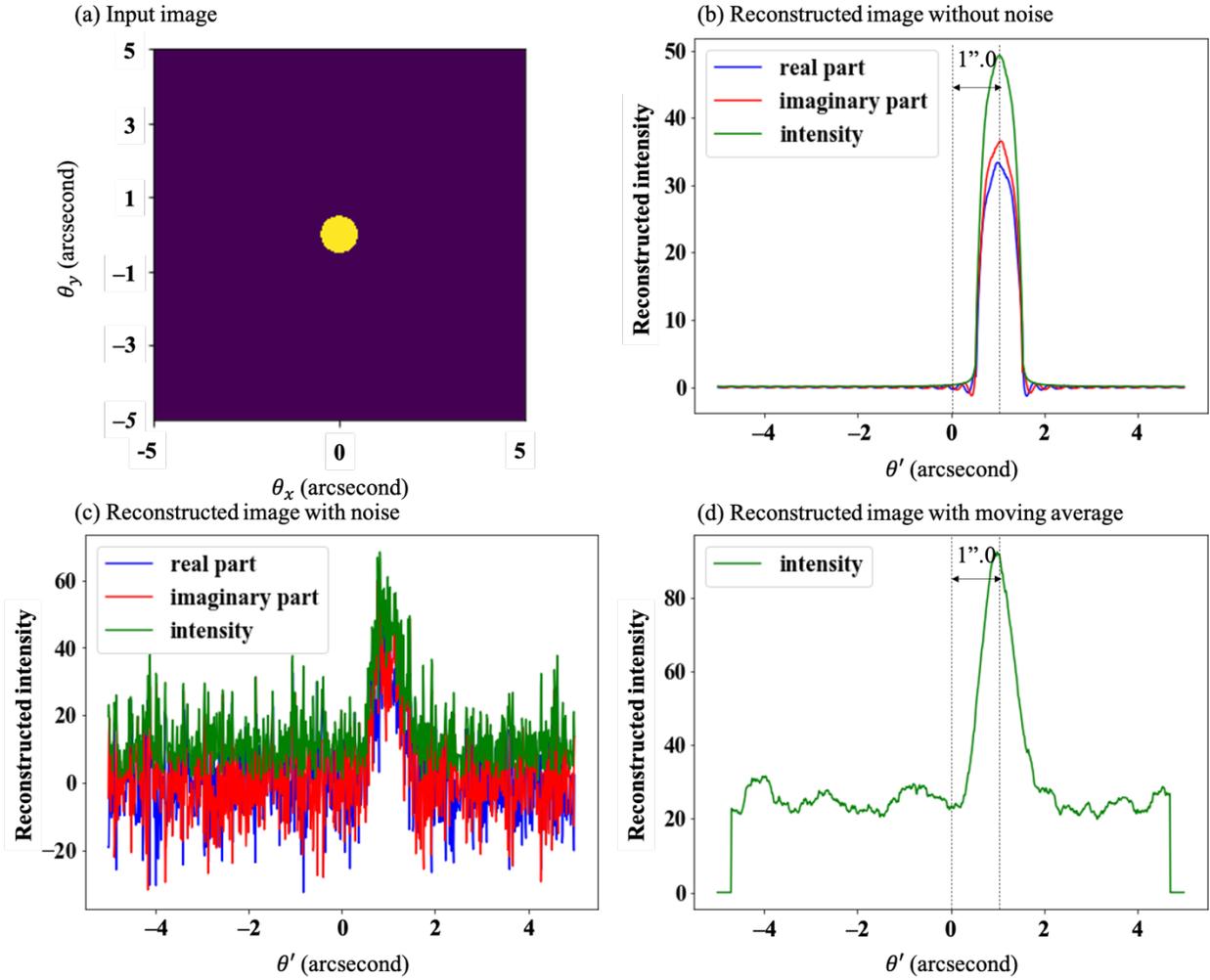

**Figure 10.** (a) Input image of the Jupiter's moon Europa in the sky. The yellow disk represents the Europa's disk with a diameter of approximately 1". (b) Reconstructed image without any observational noises, such as the shot noise and detector noise, by performing the Fourier transformation of the complex visibility taken at the short channel along the wavelength. The optical path difference between the two beams is set to 100 $\mu m$, corresponding to an offset of 1".0 on the sky. The offset value is equal to the ratio of the optical path difference to the baseline length of 20 $m$. The blue and red lines show the real and imaginary parts of the reconstructed image. The intensity indicated by the green line is the absolute value calculated from the real- and imaginary

parts. (c) Reconstructed image with the observation noises. (d) Reconstructed image with the noise reduction through the moving average.